\documentclass[10pt]{article}
\usepackage{amsmath}
\usepackage{amssymb}
\usepackage{graphicx}
\usepackage{lastpage}
\usepackage{fancyhdr}
\usepackage[pdftex, colorlinks=true, citecolor=green]{hyperref}
\usepackage{multirow}
\usepackage{lscape}
\usepackage{enumerate}
\usepackage{wrapfig}
\usepackage{lastpage}
\usepackage{times}
\usepackage{enumitem}
\usepackage{epstopdf}
\usepackage{color}
\setdescription{leftmargin=7mm,labelindent=0in}

\begin{document}

\vspace{5mm}
\noindent {\large \bf Course of the first month of the COVID 19 outbreak in the New York State counties}

\noindent \emph{Anca R\v{a}dulescu, Mathematics, SUNY New Paltz}

\vspace{3mm}
\begin{abstract}
\noindent We illustrate and study the evolution of reported infections over the month from March 1st to April 1st in the New York State as a whole, as well as in each individual county in the state. We check for exponential trends, and try to understand whether there is any correlation between the timing and dynamics of these trends and state-wide mandated measures on testing and social distancing. We conclude that the reports on April 1st may be dramatically under-representing the actual number of state-wide infections, and we propose reassessment of the data over the coming weeks using the increasing number of casualties as a validating measure. The will also allow to monitor for effects of the PAUSE directive.
\end{abstract}

\section{Introduction}

The study of early epidemic growth has historically revealed different patterns, depending on the particular pathogen, and even on the particular outbreak. While other growth patterns have also been found (e.g., polynomial), exponential growth has been a seemingly ubiquitous trend detected in data of early outbreaks of influenza, Ebola, foot-and-mouth disease, plague, measles, smallpox. That is somewhat motivated by the free spread of the pathogen in the first stages of the epidemic. Exponential growth patterns appear to also be representative for the current COVID-19 pandemic, over its first months of development in the US. In this study, we investigate these patterns specifically within the New York State, in which the epidemic has caught the widest proportions. 

Since its first confirmed cases in the US, it has become clear that the COVID-19 outbreak will globally affect all US states and territories. However, there have been essential between-state differences in the timeline and magnitude of the epidemic. It is likely that these differences are based on a variety of factors, from timing of the first contamination (earlier states were caught unprepared), to population density and social dynamics, to timing and efficiency of state-wide mandated directives for travel bans and social distancing, to timing and availability of testing. We will first illustrate such differences between three early states with most significant spread: California (9,816 infection by April 1st) , Washington State (5,588 infections) and New York State (83948 infections). Then we will focus more specifically on the New York State dynamics, trying to understand both the specific and the unifying trends in the contamination data from different counties, and to correlate these trends with state-mandated measures on testing and social distancing.

\section{Modeling methods}

We will use time series estimating the confirmed number of infections with COVID-19, as reported in the public domain by the COVID Tracking Project~\cite{testing} and by the Johns Hopkins Center for Systems Science and Engineering~\cite{COVID_archive}. In order to more easily detect potential exponential growth $N(t) = N_0 e^{\alpha t}$ (where $t$ is time in days, and $N$ is number of infected individuals), we will be considering the logarithm of the time series $\ln N = \ln N_0 + \alpha t$. This will allow to test for linearity, and -- if linear -- compute the exponential growth as the slope $\alpha$.

\begin{figure}[h!]
\begin{center}
\includegraphics[width=\textwidth]{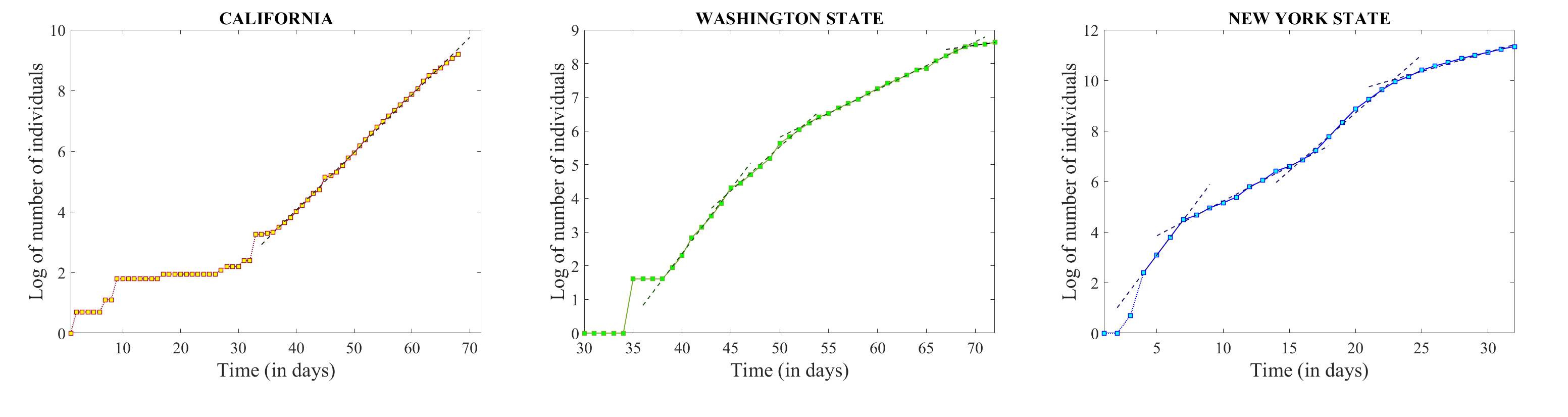}
\end{center}
\caption{\small \emph{{\bf Log time series} for the states of California (left), Washington (center) and New York (right), illustrating the portions of the time series which justify a linear fit. The best fit line is shown as a dashed curve in each case, and the values of the slope and goodness of fit are shown in Table~\ref{intro_table}.}}
\label{comparison}
\end{figure}

What one would generally expect to see in the epidemic development is an initial period with high growth rate (potentially exponential). As the growth rate subsides, the time series should move to a slower growing curve, then finally reach a peak, and eventually transition to a decreasing trend (as the epidemic is dying out). Figure~\ref{comparison} illustrates a comparison between the data for the states of California, Washington and New York. These are all states with early contamination compared to other states (January 1st in Washington, January 25th in California, March 1st in New York). The figure shows the logarithm of the raw time series, with the day of progression labeled from the start of the outbreak in the respective state. 

The unifying pattern between the three states illustrated here is that there are almost linear time windows in the evolution of each,  ultimately suggesting that all three raw time series were still growing exponentially for the time interval preceding April 1st. Table~\ref{intro_table} describes more throughly these time windows, showing the corresponding slope $\alpha$ and the goodness of fit. There are, however, significant differences in the dynamic patterns for each case -- reflected in the length, succession and slope of the linear pieces. According to this data, the number of confirmed cases in California has been on an exponential rise with slope $~ 0.18$ since day 36 (February 29th), with a slight sub-exponential tendency towards the very end. Since day 38 (February 27th) Washington has been showing a pattern which can be explained by successively moving (approximately every week) to slightly lower slope exponentials, which have flattened out to a rate close to zero by the end of March. In turn, New York State shows a sequence of oscillations in rate, with a large exponential rate over the first few days (March 4-7), even after the initial transient spike in contamination (March 1-4), followed by a reduction to a new approximately constant rate over the next 10 days (March 7-16), succeeded by another rate increase (March 16-23), and then a seeming reduction in rate (although the behavior still exhibits exponential growth, with a significant goodness of the linear fit to the data from days 23-32).

We are interested to explore whether these fluctuations are significant to the intrinsic disease dynamics, or whether they are confounded by other factors. The primary causes that would reflect in actual fluctuations of the infection rate per se are the population raised awareness to the epidemic, followed by the transportation bans and state-wide mandated closure of different venues (schools, campuses, churches, restaurants), followed by the directive to exercise social distancing. A potential confound is the limitation in testing ability, which, only indirectly related to the state-wide size of the outbreak, could lead to a significant under-representation of the real number of infections. 

While at this point there is no perfect way to discriminate between the two types of causes just based on the data, we will explore here some possibilities. One is to investigate if the trends in the infection time series synchronize with the timeline of different state-wide measures on both testing and social distancing. We work under the assumption that different New York State counties have very different social dynamics, which should reflect into a variety of patterns in the corresponding time series of county-wise confirmed infections. Along the same line, a synchronized trend among the time series of many counties is likely to be triggered by a state-wide mandated measure. Our study will therefore analyze the dynamic for each county, look for unifying trends, and attempt to interpret them in conjunction with state-wide control of these dynamics. First, we examine state wide data on testing, for the period from March 4th to April 1st, and look for trends that may underlie the patterns in confirmed infections. Efficiency of social distancing is harder to assess directly. We need to keep in mind that, while the results of a state government directed reduction in testing can be observed immediately, other factors, like social distancing measures, may need two weeks or longer to take effect.

\begin{table}[h!]
\begin{center}
\begin{tabular}{|l|c|c|c|c|}
\hline
{\bf County} & {\bf Interval} & {\bf Slope $\alpha$} & {\bf Sum of squares} & {\bf Pearson $\chi^2$}\\
\hline
California & 36-68 & 0.1898 & 0.1528 & 0.0231 \\
\hline
Washington & 38-45 & 0.3842 & 0.0181 & 0.0064\\
			& 45-52 & 0.2609 & 0.0334 & 0.0067\\
			& 52-69 & 0.1416 & 0.0146 & 0.0020\\
			& 69-72 & 0.0418 & $<10^{-4}$ & $<10^{-4}$\\
\hline
New York & 4-7 & 0.6965 & $<10^{-4}$ & $<10^{-4}$\\
		& 7-16 & 0.2739 & 0.0357 & 0.0067 \\
		& 16-23 & 0.4593 & 0.0691 & 0.0079\\
		& 23-32 & 0.1518 & 0.0347 & 0.0033\\
\hline
\end{tabular}
\end{center}
\caption{\small \emph{{\bf Rate of exponential growth} calculated as a the slope of the linear fit to the log plot. The Pearson goodness of fit statistic is also provided in each case.}}
\label{intro_table}
\end{table}

For the county-wide analysis, we use archived data on the COVID 19 outbreak evolution in the US~\cite{COVID_archive}, between March 1st (when the first case was reported in the State of New York), and April 1st (the date this study was initiated). The information on population estimates and population density in each New York county was obtained from the Department of Health web page~\cite{population}.

For the counties which estimate over 100 cases by April 1st (to permit an adequate assessment), we study during which time windows the behavior is exponential, and calculated the exponential growth rate. As before, we consider for each county the logarithm of its time series, and search for portions that could be well approximated by linear behavior. We perform a best linear fit (using the traditional least square method), and compute the slope and goodness of fit for each almost linear portion. We illustrate separately the counties which had an early infection start (as listed in the introduction) and the counties with later starts (after March 10th).

\section{Results}

\subsection{State-wide trends}

\begin{figure}[h!]
\begin{center}
\includegraphics[width=\textwidth]{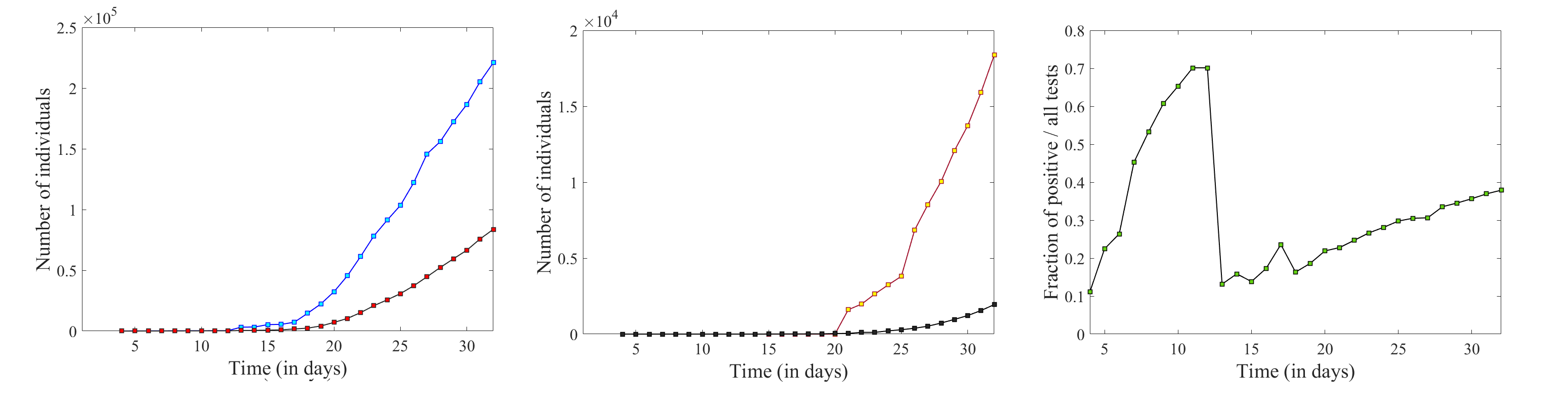}
\end{center}
\caption{\small \emph{{\bf Testing, hospitalization and casualties time series for the New York State,} over the period of March 4th to April 1st. {\bf A.} Total number of COVID-19 tests administered in blue, and number of positive tests (i.e., number of confirmed infections)  in red. {\bf B.} Number of COVID-19 diagnosed hospitalizations in yellow (data missing prior to March 20th), and number of COVID-19 associated deaths in black. {\bf C.} Fraction of confirmed positive diagnoses out of the total number of tests performed state-wide.}}
\label{NYS_combined}
\end{figure}

We performed a comparison of the evolution of the confirmed infection count versus the state-wide number of individuals tested for COVID-19, the number of COVID-associated hospitalizations and deaths. We used data from the COVID-19 Tracking repository~\cite{testing}, with raw time series represented in Figure~\ref{NYS_combined}a and b.

\begin{figure}[h!]
\begin{center}
\includegraphics[width=0.65\textwidth]{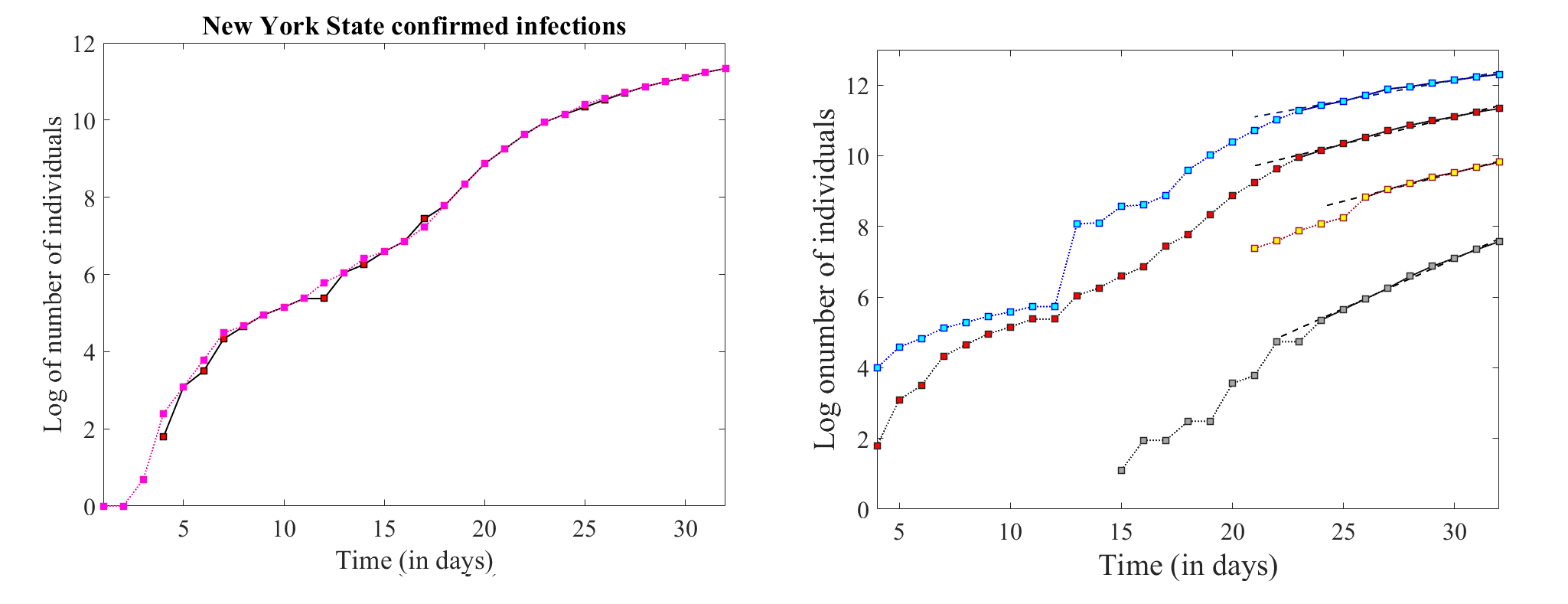}
\end{center}
\caption{\small \emph{{\bf Title.} {\bf A.} Comparison of the log time series of the confirmed infections, as provided by the data source use for this section~\cite{testing} (in red) and by the date source used in the next section~\cite{COVID_archive} (in pink). {\bf B.} Logarithmic time series corresponding to the raw time series in (A) and (B), shown with the same color coding. The illustration shows that all these functions are close to linear over a significant time period before April 1st. The slopes and goodness of fit for these time windows are shown in Table~\ref{testing_table}. }}
\label{NYS_2}
\end{figure}

In Figure~\ref{NYS_2}a, we show for cross-validation the time series of the state-wide infection count, as provided by two different sources in the public domain: the source that we used in conjunction with the county-wise infection breakdown~\cite{COVID_archive} (same as in Figure\ref{intro_table}), covering the period from March 1st to April 1st; and the source that we used in conjunction with the evolution of state-wide testing~\cite{testing}, covering the period from March 4th to April 1st. On the common portion, the two are identical, via local minor details, and exhibit the same increasing patterns at different piece-wise slopes.

\begin{table}[h!]
\begin{center}
\begin{tabular}{|l|c|c|c|c|}
\hline
{\bf Curve significance} & {\bf Interval} & {\bf Slope $\alpha$} & {\bf Sum of squares} & {\bf Pearson $\chi^2$}\\
\hline
Total tests & 23-32 & 0.1155 & 0.0236 &  0.0020\\
\hline
Positive tests & 23-32 & 0.1545 & 0.0257 & 0.0024\\
\hline
Hospitalizations & 26-32 & 0.1614 & 0.0048 & 0.0005\\
\hline
Fatalities & 24-32 & 0.2817 & 0.0193 & 0.0029\\
\hline
\end{tabular}
\end{center}
\caption{\small \emph{{\bf Rate of exponential growth} calculated as a the slope of the linear fit to the log plot. The Pearson goodness of fit statistic is also provided in each case.}}
\label{testing_table}
\end{table}

\subsection{County-wide behavior}

In the State of New York, the first signs of contagion were detected and recorded in New York City (on March 1st), then a few days later in Westchester (March 4th), Nassau (March 5th), Rockland (March 6th), Saratoga (March 7th), Suffolk and Ulster Counties (March 8th). One month later, on April 1st, the leading five counties in infection counts, with over 3,000 infected individuals, were among this list of early foci: New York City (with 47,439 reported infections), Westchester County (10,683 infections), Nassau County (9,554 infections), Suffolk County (7,605 infections), Rockland County (3,321 infections). On the other hand, Saratoga and Ulster Counties were only counting 122 and respectively 222 infections by April 1st. 

We first illustrate and analyze data in these counties with confirmed early infection (before March 10th). These are networks of communities in which the infection has already propagated for a long enough time to provide more substantial data that may allow us to understand the mechanics of this propagation. We will illustrate separately the five counties which had over 3,000 confirmed infections by April 1st, and the two counties which, despite an early start, were an order of magnitude lower in the number of infections by the same date. 

\begin{figure}[h!]
\begin{center}
\includegraphics[width=\textwidth]{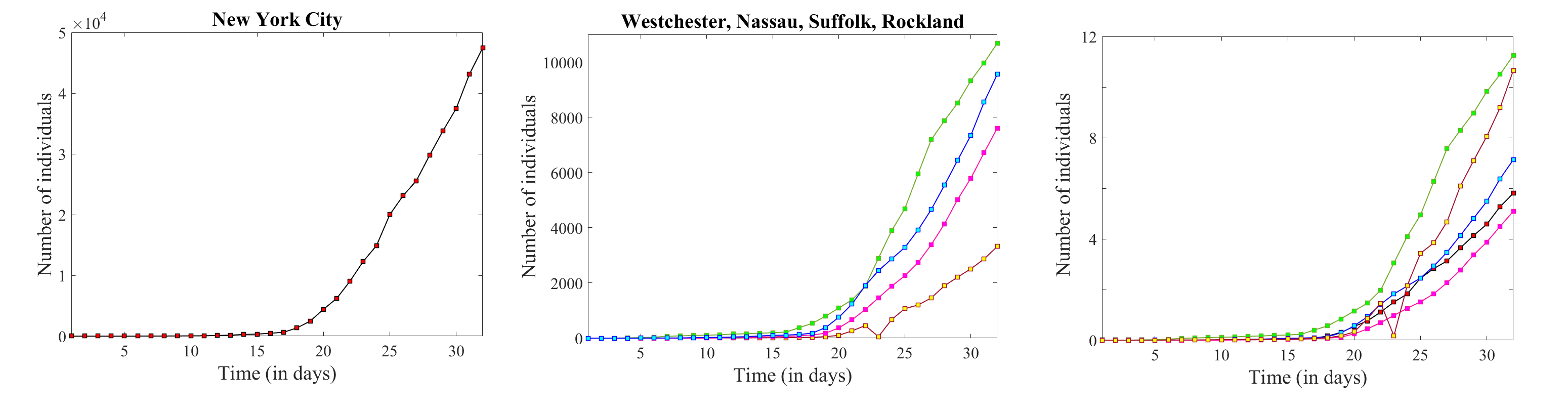}
\end{center}
\caption{\small \emph{{\bf Time series showing the progression in the county-wide number of infected individuals} up to April 1st, for the counties with early confirmed contamination (before March 10th) and over 3000 confirmed infections. {\bf A.}  New York City (red curve). {\bf B.} Westchester County (green curve), Nassau County (blue curve), Rockland County (yellow curve) and Suffolk County (purple curve). {\bf C.} Normalized time series showing the number of infected individuals per thousand of county residents, with the same color coding as in (B).}}
\label{raw1}
\end{figure}

While by April 1st New York City was transcending all other counties by a factor of at least five, and Rockland County showed the lowest numbers (Figure~\ref{raw1}a and b), in a normalized representation of infections per 1,000 individuals, New York City came fourth, following in order Westchester, Rockland and Nassau (Figure~\ref{raw1}c. It is also relatively easy to see that in different counties , confirmed infections appear to be increasing at different rates at the end of the time window: while Westchester is leading in terms of the proportion of individuals with confirmed diagnoses, Rockland is increasing at a higher rate. In order to better represent the evolution of the rate of change, and test for exponential trends, we again considered the time series in logarithmic form.

\begin{figure}[h!]
\begin{center}
\includegraphics[width=0.65\textwidth]{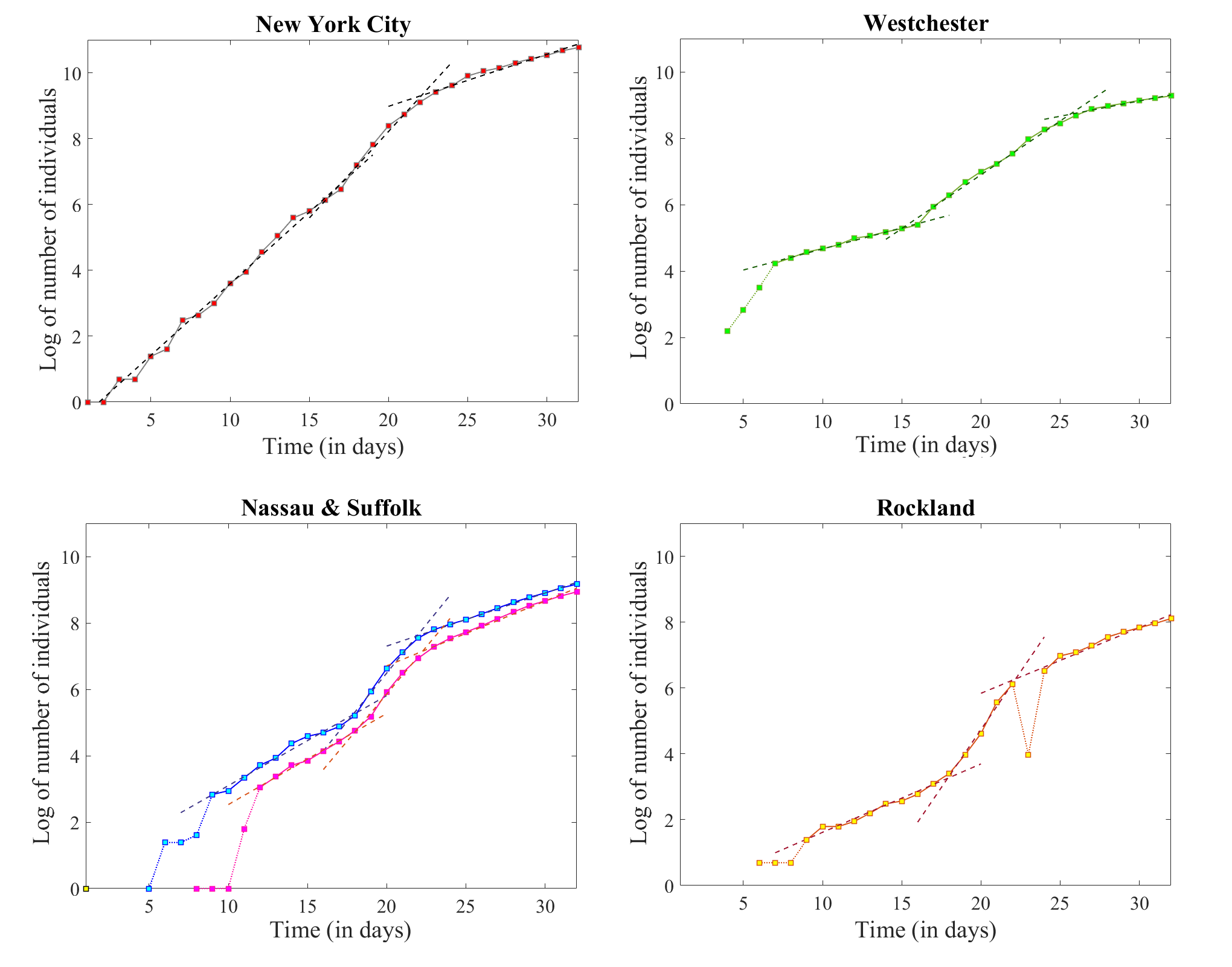}
\end{center}
\caption{\small \emph{{\bf Log time series} corresponding to the data shown in Figure~\ref{raw1}, together with the best linear fit (shown as a dashed curve in each case): for the New York City population (left panel, black curve), Westchester (green curve), Nassau (blue curve), Rockland (brown curve) and Suffolk (purple curve), in the right panel. }}
\label{log1}
\end{figure}

\begin{figure}[h!]
\begin{center}
\includegraphics[width=0.65\textwidth]{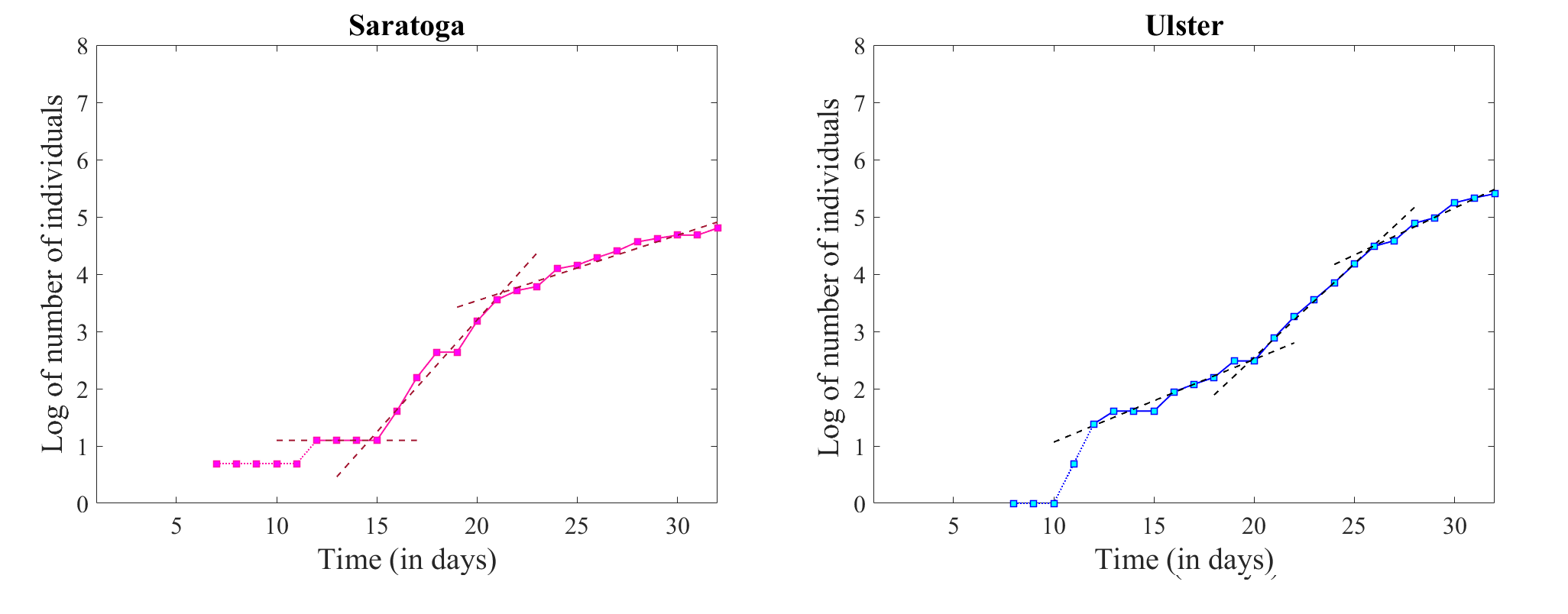}
\end{center}
\caption{\small \emph{{\bf Log time series} together with the best linear fit (shown as a dashed curve in each case): for Saratoga County (left) and Ulster County (right).}}
\label{log2}
\end{figure}

We performed linear fitting for each county, starting with the day when the first case was reported and detecting best fit linear intervals throughout the whole evolution. The slopes $\alpha$ (corresponding to the exponential growth rates) and the goodness of fit statistics are shown in Table~\ref{table1}. 

\begin{table}[h!]
\begin{center}
\begin{tabular}{|l|c|c|c|c|}
\hline
{\bf County} & {\bf Interval} & {\bf Slope $\alpha$} & {\bf Sum of squares} & {\bf Pearson $\chi^2$}\\
\hline
 New York City  & 1-17 & 0.4345 & 0.4935 & 0.0078\\
  			    & 17-22 & 0.5253 & 0.1001 & 0.0128\\
			    & 22-32 & 0.1579 & 0.0911 & 0.0093\\
\hline
Westchester  & 7-16 & 0.1269 & 0.0116 & 0.0024\\
			& 16-26 & 0.3239 & 0.1104 & 0.0165\\
			& 26-32 & 0.0922 & 0.0080 & 0.0009\\
\hline
Nassau   & 9-18 & 0.2712 & 0.1019 & 0.0256\\
		& 18-22 & 0.5853 & 0.0462 & 0.072\\
		& 22-32 & 0.1600 & 0.0123 & 0.0015\\
\hline
Suffolk   &  12-18 & 0.2744 & 0.0137 & 0.0037 \\
		& 18-22 & 0.5689 & 0.0237 & 0.0041\\
		& 22-32 & 0.1952 & 0.0543 & 0.0070\\
\hline
Rockland  & 9-18 & 0.2077 & 0.0747 & 0.0356\\
		  & 18-22 & 0.7036 & 0.0408 & 0.0088\\
		  & 22-32 & 0.1991 & 0.0923 & 0.0123\\
\hline
Saratoga & 12-15 & 0 & 0 & 0\\
		& 15-21 & 0.3910 & 0.1306 & 0.0629\\
		& 21-32 & 0.1144 & 0.0829 & 0.0194\\
\hline
Ulster     & 12-20 & 0.1444 & 0.0618 & 0.0339\\
		& 20-26 & 0.3275 & 0.0086 & 0.0029\\
		& 26-32 & 0.1641 & 0.0254 & 0.0050\\
\hline
\end{tabular}
\end{center}
\caption{\small \emph{{\bf Rate of exponential growth} calculated as a the slope of the linear fit to the log plot. The Pearson goodness of fit statistic is also provided in each case (one outlier -- March 24 -- was left out when computing the slopes for Rockland County).}}
\label{table1}
\end{table}

Notice that (after an initial transient of a few days in some cases) all these counties show a similar piece-wise linear pattern to that pointed out in the state-wide logarithmic time series. In each case, we were able to identify three pieces: a milder increasing segment ending between March 16-18, followed by a steeper segment ending on March 22 (26 in the case of Westchester), followed again by a linear segment with more relaxed slope, and very significant goodness of fit.

While the numbers were an order of magnitude lower for Saratoga and Ulster Counties, the same three-piece pattern was surprisingly apparent in their time series as well, suggesting that the piece-wise linear effect with alternating steepness which we observed in the state wide dynamics was not an average effect, but is rather present in every one of the counties with early contamination. This suggests that the effect is either a reflection of specific New York social dynamics across counties -- leading to a subsequently specific spread pattern and speed -- or more likely the effect of state-wide unifying controls. 

We want to further investigate whether the transitions between linear pieces use as temporal reference the time of the original confirmed infection in the corresponding county, or if they are synchronized at a central level. Since the counties represented so far had similar times of original infection, we need to compare with the similar data in counties with a different timeline.

\begin{figure}[h!]
\begin{center}
\includegraphics[width=\textwidth]{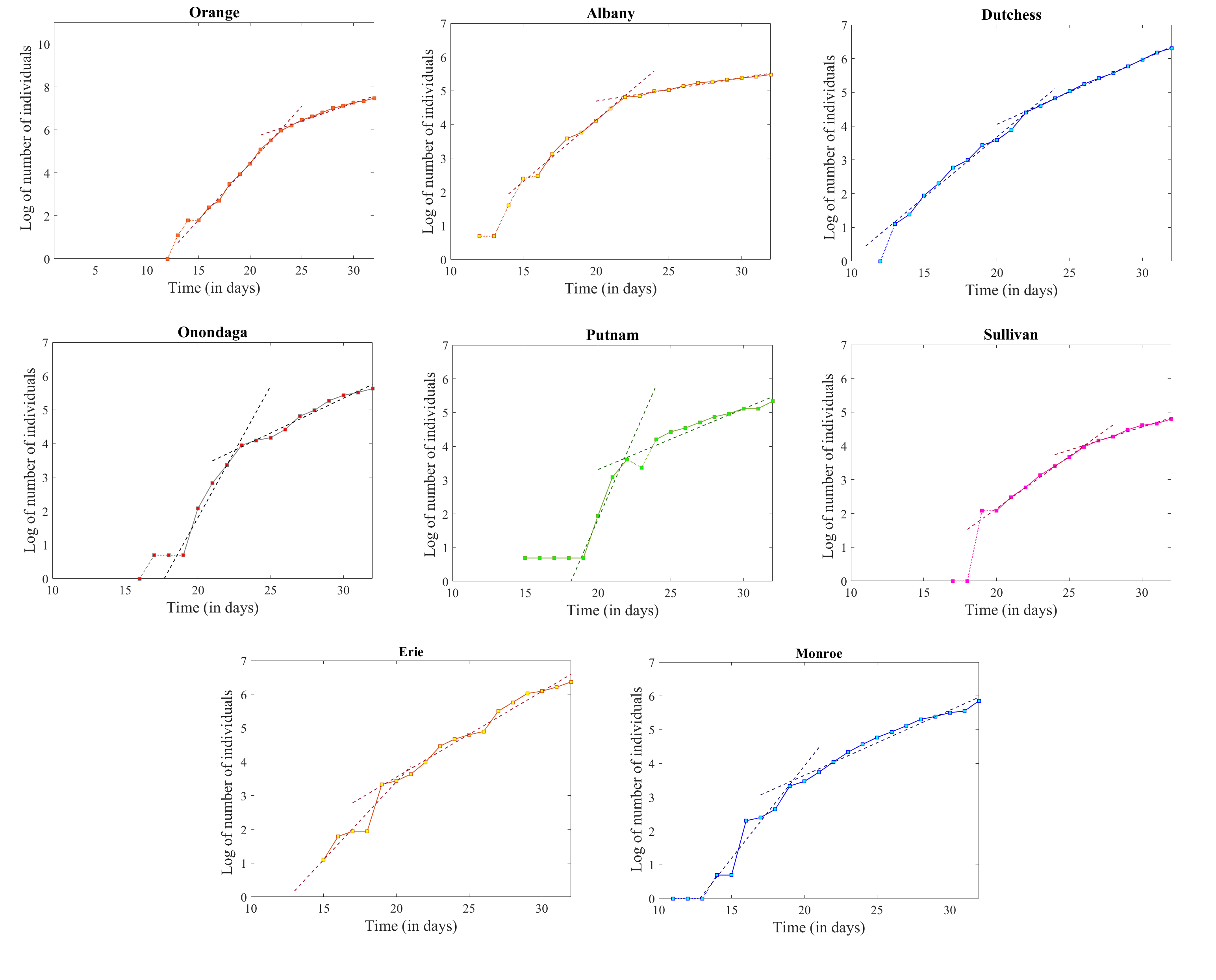}
\end{center}
\caption{\small \emph{{\bf Log time series} corresponding to the data shown in Figure~\ref{raw1}, together with the best linear fit (shown as a dashed curve in each case): for the New York City population (left panel, black curve), Westchester (green curve), Nassau (blue curve), Rockland (brown curve) and Suffolk (purple curve), in the right panel. }}
\label{log3}
\end{figure}

The only other nine New York State counties which had a count of over 100 infection cases by April 1st are as follows, in order of the number: Orange (1,756), Erie (582), Dutchess (547), Monroe (349), Onondaga (277), Albany (240), Ulster (222), Putnam (207), Saratoga (122) and Sullivan (121). Of these, as specified before, Saratoga and Ulster had confirmed cases by March 10th; the other counties have later contamination dates. Their logarithmic time series, together with the linear fits, are shown in Figure~\ref{log3}; the slopes and goodness of fit statistics are described in Table~\ref{table3}. The time series of these counties all exhibit only two linear segments (the first with steeper slope, followed by one with flatter slope) with the transition occurring over the same short time window (March 22-26) as the similar transition in the counties with early starts. This confirms the idea that the evolution in this later batch is not replicating the evolution of the earlier counties from the very start, but rather is correlating with their current evolution (further suggesting that the reason behind these trends is centralized control).

\begin{table}[h!]
\begin{center}
\begin{tabular}{|l|c|c|c|c|}
\hline
{\bf County} & {\bf Interval} & {\bf Slope $\alpha$} & {\bf Sum of squares} & {\bf Pearson $\chi^2$}\\
\hline
Orange  & 15-23 & 0.5293 & 0.0527 & 0.0155\\
		& 23-32 & 0.1655 & 0.0499 & 0.0074\\
\hline
Albany & 16-22 & 0.3640 & 0.0810 & 0.0268\\
 		& 22-32 & 0.0688 & 0.0116 & 0.0023\\
\hline
Dutchess & 13-22 & 0.3583 & 0.1050 & 0.0439\\
		& 22-32 & 0.1913 & 0.0073 & 0.0014\\
\hline 
Onondaga & 20-23 & 0.6149 & 0.0108 & 0.0042\\
		  & 23-32 & 0.2061 & 0.0850 & 0.0174\\
\hline
Putnam & 19-22 & 0.9898 & 0.1477 & 0.0707\\
		& 22-32 & 0.1566 & 0.1164 & 0.0271\\
\hline
Sullivan & 20-26 & 0.3092 & 0.0113 & 0.0042\\
		& 26-32 & 0.1366 & 0.0105 & 0.0024\\
\hline
Erie & 15-19 & 0.4621 & 0.4982 & 0.2044\\
      & 19-32 & 0.2537 & 0.2827 & 0.0553\\
\hline
Monroe & 14-19 & 0.5465 & 0.6212 & 0.4150\\
		&19-32 & 0.1928  & 0.2246 & 0.0490\\
\hline
\end{tabular}
\end{center}
\caption{\small \emph{{\bf Rate of exponential growth} calculated as a the slope of the linear fit to the log plot. The Pearson goodness of fit statistic is also provided in each case.}}
\label{table3}
\end{table}

\section{Discussion}

In this study, we focused on the dynamics of the COVID-19 epidemic in the state of New York for the first month of the outbreak (March 1st to April 1st), and we analyzed state-wide data on confirmed infections, testing, hospitalizations and deaths, as well as county-wide data on infection rates, for the counties in the state that were reporting over 100 infections by April 1st. Our primary goal was to determine whether it is possible to dissociate between the real, infection-related effects of social distancing measures, and the reporting effects produced by the limitations in testing. In particular, we identified a signature in the evolution of reported infections, specific to the state on New York, characterized by following exponential growths with rates oscillating between higher and lower values, for different time windows, eventually settling to a relatively low exponential rate. We aimed understand the factors behind the transitions, and whether the trend shown at the start of April corresponds to a damping in infections, or is an artifact.

When looking at the state-wide patterns of testing, we observed a substantial spike between March 12th and March 13th. While this is not obvious in the raw data in Figure~\ref{NYS_combined}a, it is clear in the logarithmic plot in Figure~\ref{NYS_2}b, as well as in the sudden drop in the fraction of positive tests shown in Figure~\ref{NYS_combined}c (higher availability of tests allowed for tests on more susceptible individuals who were not necessarily infected). In fact, notice that before March 12th, this ratio had steadily increased to the point where over 70\% of tests were positive; on March 13th, this ratio dropped to under 15\%, but promptly started creeping up again, having climbed back to almost 40\% by April 1st. This is likely the promoter for the ramping up in the slope of detected infections (a comparison between the blue and red curves in Figure~\ref{NYS_2} suggests that the two effects occurred simultaneously at the overall state level). For specific counties, the distribution of this transition point in the slope ranged from March 12 to March 18, but can still be viewed as a potential effect of the increase in testing (especially since the counties with later infection onset do not show this transition, and start off with a segment of higher exponential rate directly). In this light, these slopes are expected to show a more faithful (if not perfect) representation of the actual growth in infections, since it is less restricted by testing limitations.

Interestingly enough, there does not seem to be any disruption to this existing trend in positive versus total number of tests around March 20th, in conjunction with the state government directive to refocus testing primarily around individuals with high hospitalization risk. However, this date coincides with the last transition in the infection curve for New York State, from a higher exponential diagnosis rate $\alpha \sim 0.46$ to a lower rate $\alpha \sim 0.15$. In support of this observation is that all counties exhibit a similar transition at dates distributed between March 19 and March 23, from a higher exponential rate (with mean $\mu_\alpha = 0.51$ and standard deviation $\sigma_\alpha = 0.18$ between the 15 counties examined) to a milder rate (with mean $\mu_\alpha = 0.16$ and standard deviation $\sigma_\alpha = 0.04$). This interval appears too early to reflect any of the state mandated initial closures (which were initiated around March 12th, would need at least two weeks to take effect and would likely not be as tightly synchronized between counties). This is also too early to reflect the PAUSE directive, which could  produce a highly synchronized response between counties, but the effects of which are not expected to be immediate.

A strong hypothesis to consider is that the number of infections (with a rate of exponential increase possibly even transcending the reported $\alpha \sim 0.5$ up to March 20th) overtook the sampling rate provided by the testing. In order to maintain realistic reporting, testing availability would have had to also increase at a comparable pace. The state governor directive on March 20th allegedly had as an effect restricting testing on potentially positive and negative COVID-19 diagnoses at the same extent (people being instructed to refrain from testing unless in very serious condition). While there was no visible cut in testing, or additional imbalance in positive versus negative tests, the directive may have effectively capped the reported infections to the point  where the rate of the new infections ($\alpha =0.15$) settled only a little higher than the overall rate of testing ($\alpha = 0.115$) after March 20th. 

Let's assume here that the lowering of the exponential rate is only a testing-driven artifact, and the infections were still occurring up to April 1st at the same, higher rate $\alpha = 0.46$ shown prior to the transition in slope on March 23rd. Then the actual number of infections in New York State would be an estimate of 1,302,800 infected individuals by April 1st, compared to the reported 83,712, and the projected 80,524 when using the $\alpha = 0.15$ lower rate. Additional information will contribute over the next days and weeks to assessing whether this is the case or not. One option could be provided by monitoring whether the changes expected to follow PAUSE will actually appear in the infection time series. Another, potentially more reliable assessment will come from the evolution of casualties, evolving prior to April 1st at an exponential rate $\alpha = 0.28$ higher than the infection rate estimated from the confirmed cases, and which is expected to increase (assuming that COVID-19 susceptible deaths are actually tested).

When considering the number of confirmed infections for each county on April 1st, one simple observation is that they did not correlate with the county total population, but rather with the population density, as reported in each county by the NYS Health Department (correlation coefficient $R=0.4195$, significance value $p=0.0007$). This suggests that these numbers may still reflect, at least to some extent, the type of social dynamics associated to the population density distribution in each county. Our future work involves  quantifying county specific evolution of social dynamics (from how it reflected into travel patterns to different destinations within the community), and understanding whether these dynamics contributed to the slopes for each county, and to the transitions between regimes.


\bibliographystyle{plain}

\end{document}